\documentclass{SciPost}

\hypersetup{
    colorlinks,
    linkcolor={red!50!black},
    citecolor={blue!50!black},
    urlcolor={blue!80!black}
}

\usepackage[bitstream-charter]{mathdesign}
\usepackage{comment}

\DeclareSymbolFont{usualmathcal}{OMS}{cmsy}{m}{n}
\DeclareSymbolFontAlphabet{\mathcal}{usualmathcal}

\fancypagestyle{SPstyle}{
\fancyhf{}
\lhead{\colorbox{scipostblue}{\bf \color{white} ~SciPost Physics Proceedings }}
\rhead{{\bf \color{scipostdeepblue} ~Submission }}

\fancyfoot[C]{\textbf{\thepage}}
}

\newcommand{\eq}[1]{\eqref{eq:#1}}
\newcommand{\fig}[1]{Fig.~\ref{fig:#1}}

\begin{document}

\pagestyle{SPstyle}

\begin{center}{\Large \textbf{\color{scipostdeepblue}{
Stability of the Higgs Potential in the Standard Model and Beyond
}}}\end{center}

\begin{center}\textbf{
Tom Steudtner\textsuperscript{1$\star$}
}\end{center}

\begin{center}
{\bf 1} TU Dortmund University, Department of Physics, Otto-Hahn-Str.~4, D-44221~Dortmund, Germany
\\[\baselineskip]
$\star$ \href{mailto:email1}{\small tom2.steudtner@tu-dortmund.de}
\end{center}

\definecolor{palegray}{gray}{0.95}
\begin{center}
\colorbox{palegray}{
  \begin{tabular}{rr}
  \begin{minipage}{0.36\textwidth}
    \includegraphics[width=60mm,height=1.5cm]{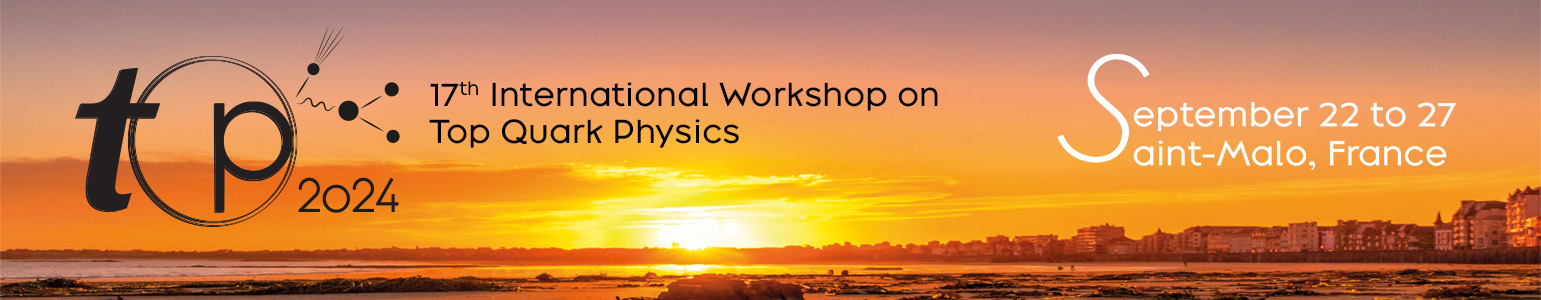}
  \end{minipage}
  &
  \begin{minipage}{0.55\textwidth}
    \begin{center} \hspace{5pt}
    {\it The 17th International Workshop on\\ Top Quark Physics (TOP2024)} \\
    {\it Saint-Malo, France, 22-27 September 2024
    }
    \doi{10.21468/SciPostPhysProc.?}\\
    \end{center}
  \end{minipage}
\end{tabular}
}
\end{center}

\section*{\color{scipostdeepblue}{Abstract}}
\textbf{\boldmath{%
The question of stability of the Higgs potential in the Standard Model is revisited employing advanced theoretical precision and recent experimental results. We show that the top mass and strong coupling constants are key observables in order to reach or refute absolute stability. We highlight new physics scenarios that lead to a decisive stabilisation of the Higgs sector. These proceedings summarise findings first reported in~\cite{Hiller:2022rla,Hiller:2024zjp}.
}}

\vspace{\baselineskip}

\noindent\textcolor{white!90!black}{%
\fbox{\parbox{0.975\linewidth}{%
\textcolor{white!40!black}{\begin{tabular}{lr}%
  \begin{minipage}{0.6\textwidth}%
    {\small Copyright attribution to authors. \newline
    This work is a submission to SciPost Phys. Proc. \newline
    License information to appear upon publication. \newline
    Publication information to appear upon publication.}
  \end{minipage} & \begin{minipage}{0.4\textwidth}
    {\small Received Date \newline Accepted Date \newline Published Date}%
  \end{minipage}
\end{tabular}}
}}
}


\vspace{10pt}
\noindent\rule{\textwidth}{1pt}
\tableofcontents
\noindent\rule{\textwidth}{1pt}
\vspace{10pt}


\section{Introduction}
\label{sec:intro}
\begin{figure}
  \centering
  \includegraphics[scale=.6]{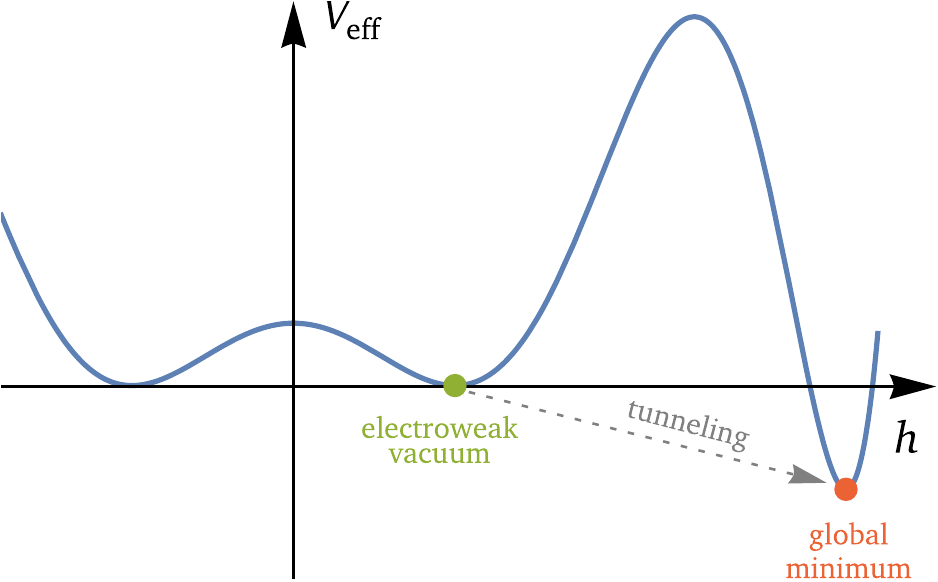}
  \caption{Schematic depiction of the effective potential $V_\text{eff}$ as a function of the Higgs field value $h$ in the SM. The electroweak vacuum $h=v$ is marked in green, a global minimum $h \gg v$ is shown in orange.}
  \label{fig:Veff-schematic}
\end{figure}

Ever since the discovery of the Higgs boson, evidence persists that the Higgs potential, schematically depicted in \fig{Veff-schematic}, is metastable. The electroweak vacuum is not the true ground state, but a deeper minimum of the potential exists, see e.g.~\cite{Hiller:2024zjp,Masina:2024ybn} for recent works. However, the tunneling rate into the global minimum is extremely low, such that  the lifetime of the electroweak is much larger than the age of the universe~\cite{Andreassen:2017rzq,Chigusa:2018uuj}.
 Thus, the metastability of the Standard Model is not necessarily a problem. Nevertheless, the question arises why the Standard Model is in such a fine-tuned proximity to absolute stability. In particular, it is intriguing if higher precision in experiment and theory shift this fate more towards stability, or, if metastability prevails.  
Moreover, the metastability situation could be understood as handle on new physics. Extensions of the SM have various motivations and could be settled anywhere between the electroweak and the Planck scale. The stability of the Higgs potential can be utilised to constrain the parameter space of SM extensions. 
In fact, stabilisation of the Higgs potential can even be promoted to the primary paradigm of new physics model building, and viable SM extensions derived from it~\cite{Hiller:2022rla,Hiller:2024zjp}.

At any rate, each scenario builds upon the question if absolute stability in the SM is still within a credible uncertainty of experimental inputs and theoretical predictions, or if absolute stability can be excluded. This will be addressed in Sec.~\ref{sec:Determination} and Sec.~\ref{sec:Discussion}. BSM scenarios that stabilise the SM are discussed in Sec.~\ref{sec:BSM}. 

\section{Effective Potential}\label{sec:Determination}
There are several stages to computing the stability of the SM Higgs potential:
\begin{enumerate}
  \item Measure observables to fix all SM parameters.
  \item Translate these observables into running $\overline{\text{MS}}$ couplings.
  \item Compute the effective potential and investigate its minima.
  \item If necessary, compute the lifetime for approximately degenerate minima.
\end{enumerate}

The first step uses the measured pole masses of the Higgs $M_h$, top quark $M_t$, $Z$ boson $M_Z$ and charged leptons $M_{e,\mu,\tau}$, the $\overline{\text{MS}}$ masses of the light quarks at various renormalisation scales $m_b(m_b)$, $m_c(m_c)$, $m_{u,d,s}(2\,\text{GeV})$, 
the 5-flavour strong coupling at the $Z$ pole scale  $\alpha_s^{(5)}(M_Z)$, the fine structure constant $\alpha_e$
and its hadronic threshold correction $\Delta\alpha_e^{(5)}(M_Z)$ as well as the Fermi constant $G_F$. 
Central values and uncertainties of these observables are taken from the 2024 edition of the PDG~\cite{ParticleDataGroup:2024cfk}.

These observables are related to $\overline{\text{MS}}$ couplings employing the fit~\cite{Alam:2022cdv}, which utilises high-loop thresholds corrections and resummations, see \cite{Martin:2019lqd} for an overview. We obtain the SM couplings 
\begin{equation}
  \alpha_{1,2,3} = \frac{g_{1,2,3}^2}{(4\pi)^2}\,, \qquad \alpha_{f} = \frac{y_f^2}{(4\pi)^2}\,, \qquad \alpha_\lambda = \frac{\lambda}{(4\pi)^2}
\end{equation}
at a reference scale $\mu_\text{ref} = 200~\text{GeV}$. Here, $f$ counts over all fermions flavours.

A second minimum of the effective potential may emerge at large field values $h \gg v$.  Thus, only the terms quartic in the field values are interesting for the analysis. 
A suitable ansatz at the reference scale $ \mu = \mu_\text{ref}$ is 
\begin{equation}\label{eq:Veff}
  V_\text{eff} = \frac{1}{4} \lambda_\text{eff}(h) \, e^{4 \bar{\Gamma}(h)} h^4\,.
\end{equation}
Absolute stability requires that there is no minimum deeper than the electroweak one, i.e. avoiding the situations shown in \fig{Veff-schematic}.
In good approximation, this is the case when 
\begin{equation}
  \lambda_\text{eff}(h) > 0 \quad \text{ for all field values $h \gg v$, }
\end{equation}
as all other factors in \eq{Veff} are manifestly positive.
$\lambda_\text{eff}(h)$ can be computed to high loop orders, combining fixed-order calculations of the effective potential with resummations of all large logarithms $\log h/\mu_\text{ref}$. Details of this procedure and all relevant literature results are found in~\cite{Hiller:2024zjp}.

If $\lambda_\text{eff}(h)$ becomes only slightly negative, the potential might be metastable as the lifetime of the electroweak minimum before tunneling into the true vacuum is compatible with the observed age of the universe. To estimate this, the decay rate between both vacua has to be computed. As we are only interested in absolute stability, this step is omitted here.

\section{Stability Analysis}\label{sec:Discussion}

Now, we evaluate the impact of the experimental data on the stability of the Higgs potential. The focus will be on observables which can push the SM into a regime of absolute stability within the $5\sigma$ uncertainty around their central values. As only the couplings $\alpha_{1,2,3,t,\lambda}(\mu_\text{ref})$ are sizable enough to yield non-negligible contributions to the effective potential, the lighter quark and lepton masses are irrelevant for this discussion. Furthermore, the electroweak observables determining $\alpha_{1,2}$ have such a small uncertainty that absolute stability cannot be reached anywhere near the $5\sigma$ uncertainty. We are left with the couplings $\alpha_{\lambda,t,3}$ and their corresponding observables: the Higgs and top pole masses and the strong coupling constant. The PDG values~\cite{ParticleDataGroup:2024cfk}, as well as their shift required to stabilise the Higgs potential are listed in Tab.~\ref{tab:ParticleDataGroup:2024cfk}.
\begin{table}[ht!]
  \centering
  \begin{tabular}{|ll|rr|rr|}
       \hline
       \multicolumn{2}{|c|}{PDG Observables~\cite{ParticleDataGroup:2024cfk}}&  \multicolumn{2}{|c|}{Absolute Stability}\\ 
       \hline 
      $M_h/\text{GeV}$&$=  125.20(11)$ & $>127.85$ &$+24.0\sigma$ \\
      $\alpha_s^{(5)}(M_Z) $ & $=  0.1180(9)$ & $>0.1213$ & $+\phantom{0}3.7 \sigma$ \\
      $M_t^\sigma/\text{GeV}$ & $= 172.4(7)$ & $<171.10$ & $- \phantom{0}1.9 \sigma$ \\
       $M_t^\text{MC}/\text{GeV}$ &$=  172.57(29)$  & & $- \phantom{0}5.1 \sigma$ \\
       \hline
  \end{tabular}
  \caption{
  Strong coupling, Higgs or top mass and their uncertainty from PDG 2024~\cite{ParticleDataGroup:2024cfk} edition. For each observable, the shift around the central values  required to stabilize the effective potential before the Planck scale is given. Top masses from cross-section measurements $(M_t^\sigma)$ and Monte-Carlo generators $(M_t^\text{MC})$ are distinguished. Table adopted from~\cite{Hiller:2024zjp}.
  }
  \label{tab:ParticleDataGroup:2024cfk}
\end{table}

\begin{figure}
  \centering
  \includegraphics[scale=.7]{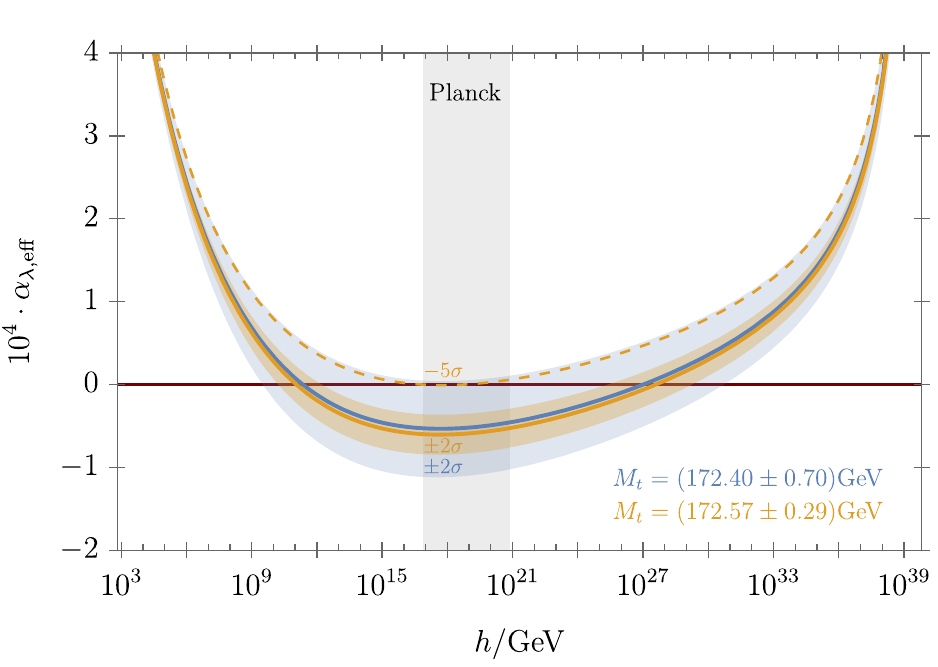}
  \caption{Effective quartic interaction $\alpha_{\lambda,\text{eff}}(h) \equiv\lambda_\text{eff}(h)/ (4\pi)^2$ of the quantum-improved potential   
  as a function of the Higgs field $h$. 
  Several choices for the top pole mass are shown, including the PDG central value of $M_t^\sigma$ (solid blue) and $M_t^\text{MC}$ (solid orange) as well as their respective $\pm2\sigma$ uncertainty (shaded blue and orange), and the $-5\sigma$ value of $M_t^\text{MC}$ (dashed orange).
  Other observables are taken as PDG central values~\cite{ParticleDataGroup:2024cfk}. 
  Only potentials that remain above the red line at $\alpha_{\lambda,\text{eff}} = 0$ for all values of $h$ are stable. 
  }
  \label{fig:Veff-Mt}
\end{figure}

While the Higgs pole mass is central to extract $\alpha_\lambda(\mu_\text{ref})$, the impact on stability is meager. The observable has a small uncertainty and gives rise to a coupling $\alpha_\lambda < \alpha_{3,t}$ already small at the reference scale and almost irrelevant around the Planckian regime. The RG running, and in turn the resummation of large field values is the driving factor behind the formation of a second minimum, where each of the couplings $\alpha_{1,2,3,t}$ is more dominant.
As such, an upwards shift of the strong coupling by $3.7\sigma$ would increase $\alpha_3$ sufficiently to stabilise the Higgs potential. Similarly, a smaller top mass and the corresponding decrease of $\alpha_t$ has the same effect.
The PDG curates two values for the top mass, one extracted from cross section measurements $M_t^\sigma$, and one from Monte Carlo template fits, $M_t^\text{MC}$. 
Both have compatible central values, but their uncertainties are vastly different. 
$M_t^\sigma = 172.4(7)~\text{GeV}$ is quoted with a more conservative uncertainty, which puts stability only $1.9\sigma$ below the central value. On the other hand, $M_t^\text{MC}=172.57(29)~\text{GeV}$ excludes absolute stability as a $5.1 \sigma $ downward shift would be required. 
The effective potential for both choices of top masses and their uncertainties is displayed in \fig{Veff-Mt}.
Alternatively, a recent combination of ATLAS and CMS data from run 1 at the LHC obtains a Monte Carlo mass $M_t^\text{MC,LHC} = 172.52(33)~\text{GeV}$~\cite{ATLAS:2024dxp}, which is merely $-4.3\sigma$ away from stability. 
Note that the Monte-Carlo mass $M_t^\text{MC}$ is not identical to the top pole mass $M_t$. The difference is due to modelling assumptions of the Monte-Carlo generators and not quantitatively understood, though~\cite{Hoang:2020iah} estimates $|M_t^\text{MC} - M_t| \approx 0.5~\text{GeV}$, which should be included on top of the uncertainties quoted on $M_t^\text{MC}$. Moreover, the relation between pole and $\overline{\text{MS}}$ mass includes non-perturbative corrections encoded in the renormalon ambiguity. Thus, a conservative estimate should also add a theory uncertainty of order $\Lambda_\text{QCD}$ to the $\overline{\text{MS}}$ mass. Circumventing the pole mass and directly reconstructing the $\overline{\text{MS}}$ mass from experimental data would eliminate this uncertainty.

\begin{figure}
  \centering
  \includegraphics[scale=.7]{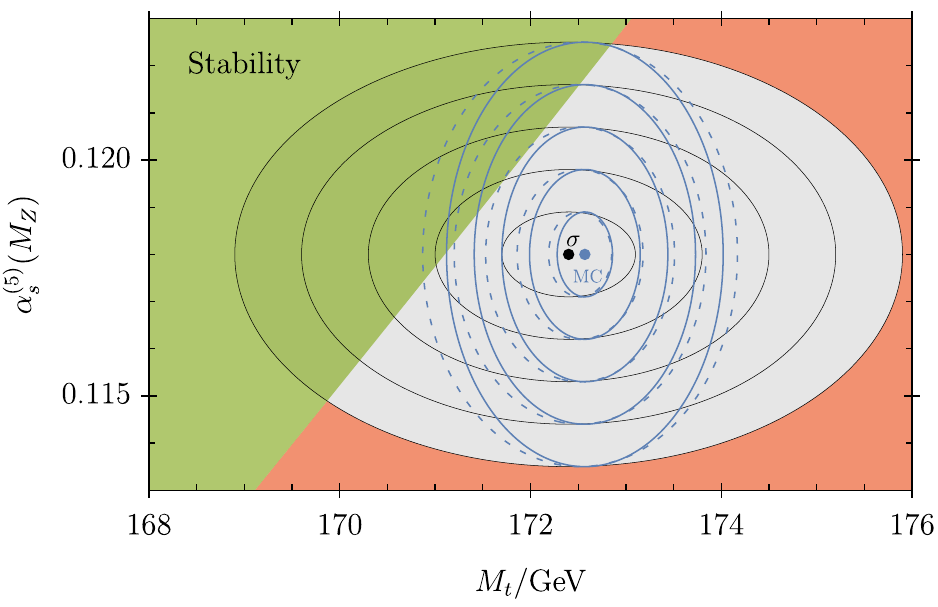}
  \caption{ 
    Stability region for the SM Higgs effective potential as a function of the top pole mass $M_t$ and the strong coupling constant $\alpha_s^{(5)}(M_Z)$. 
    An absolutely stable potential is marked in green, while red signifies meta- and rapid instability. Central values from the PDG~\cite{ParticleDataGroup:2024cfk} are marked with a dot, with $1\sigma$ uncertainties of the observables added in quadrature are denoted as rings. We distinguish the top mass $M_t^\sigma$ from cross-section measurements (black) and the Monte Carlo mass $M_t^\text{MC}$ (blue). Uncertainty rings for the Monte Carlo mass uncertainty~\cite{ATLAS:2024dxp} are drawn as dashed blue lines.
  }
  \label{fig:region-Mt-as}
\end{figure}

\begin{figure}
  \centering
  \includegraphics[scale=.7]{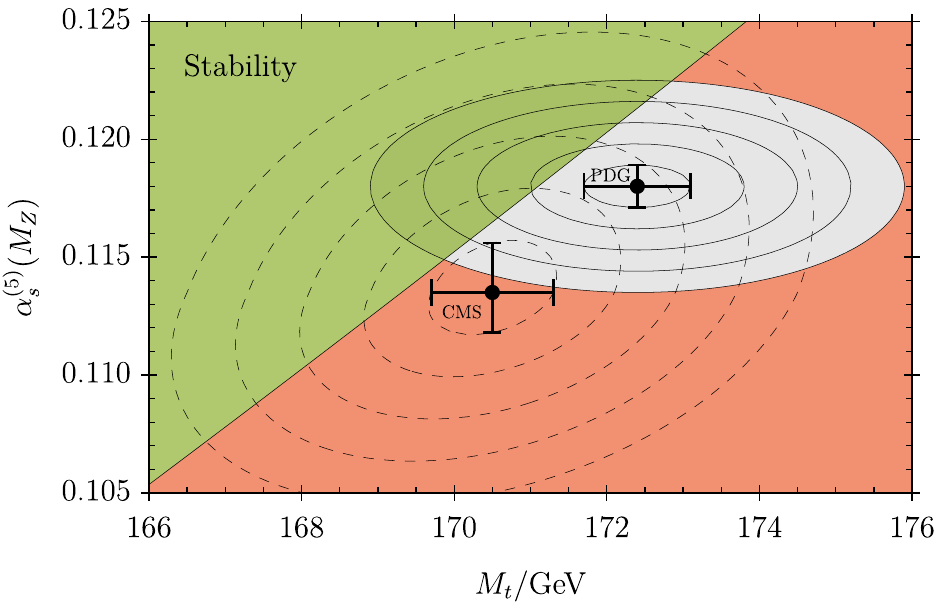}
  \caption{ 
    Stability region (green) for the SM Higgs effective potential as a function of the top pole mass $M_t$ and the strong coupling constant $\alpha_s^{(5)}(M_Z)$.  PDG central values (using $M_t^\sigma$)~\cite{ParticleDataGroup:2024cfk} with uncorrelated $1\sigma$ uncertainties added in quadrature (solid rings) are compared with the CMS study~\cite{CMS:2019esx} (dashed uncertainty regions) with a correlation $\rho = 0.31$. Figure adopted from~\cite{Hiller:2024zjp}. }
  \label{fig:region-Mt-as-2}
\end{figure}

Overall, vacuum stability is placed at the mercy of precision measurements of the top mass and strong coupling. \fig{region-Mt-as} indicates how far the SM is away from absolute stability in terms of these observables when using PDG averages and uncertainties~\cite{ParticleDataGroup:2024cfk}. Using the more conservative value $M_t^\sigma$, the top mass is the largest source of uncertainty and stability might be within the region of combined $2\sigma$ uncertainties of each observable, 
neglecting correlations between them.
If $M_t^\text{MC}$ is taken at face value, the uncertainty of $\alpha_s$ becomes more dominant, and stability might be within the region of combined $4\sigma$ uncertainties.
Neither scenario allows to exclude absolute stability at a $5\sigma$ significance. On the contrary, the distance to the stability region in terms of the combined uncertainty might be underestimated as correlation between both observables is neglected. A counter-example the older CMS work~\cite{CMS:2019esx}, where the correlation has been taken into account. As depicted in \fig{region-Mt-as-2}, the uncertainty is pushed away from the stability region compared to the uncorrelated case. Thus, keeping track of correlations is vital in order to refute stability at $5\sigma$.

\section{New Physics Portals}\label{sec:BSM}

There are three approaches how to stabilise already the tree-level Higgs potential with minimal extensions of the SM: the \textit{gauge-}, \textit{yukawa-} and \textit{scalar portal mechanism}~(see \cite{Hiller:2022rla,Hiller:2024zjp} and references therein). 
They are minimal as only one type of new physics field needs to be introduced and the SM gauge group is maintained. Each one of these mechanisms predicts realistic new physics models that can be searched for at current and future colliders. 

All portals rely on the stabilisation of the RG flow of the Higgs quartic $\alpha_\lambda(\mu)$. The gauge portal achieves this with charged BSM fields which, mediated by gauge interactions, stabilise the Higgs at loop level. In principle any new physics fields will do, though BSM scalars come with their own stability sector intertwining with the Higgs. On the other hand, BSM fermions may not couple to the Higgs directly. Depending on their representations, BSM leptons lift the RG evolution of $\alpha_1(\mu)$ and/or $\alpha_2(\mu)$, which in turn lifts the running of $\alpha_\lambda(\mu)$ as a two-loop effect. In addition, BSM quarks cause a generally larger coupling $\alpha_3(\mu)$, which accelerates the decrease of the top Yukawa $\alpha_t(\mu)$ towards higher scales. This as well results in an uplift of $\alpha_\lambda(\mu)$ now at three-loop level, though this strong-portal may be just as sizable as the electroweak portal. 
In either case, there are constraints on the mass scale, quantity and representations of new physics fields which need to be sufficient to stabilise the running of $\alpha_\lambda(\mu)$ but without introducing sub-planckian Landau poles. A more detailed discussion is found in \cite{Hiller:2022rla}.

Yukawa portals open when the Higgs is coupled to BSM fermions via new Yukawa interactions. At small coupling values, these Yukawa portals act destabilising and are in competition with gauge portals. At more sizable values, these Yukawas may trigger walking regimes in the RG flow, where the running of $\alpha_\lambda$ is significantly slowed by the vicinity to pseudo-fixed points for a wide range of scales, avoiding instabilities~\cite{Hiller:2022rla}. 

Scalar portals rely on a BSM scalar $S$ in addition to the SM Higgs doublet $H$. Both scalar sectors are intertwined via a portal interaction $\delta\,(H^\dagger H) (S^\dagger S)$. The running of the Higgs quartic receives an uplift by the manifestly positive contribution $\beta_\lambda = \beta_\lambda^\text{SM} + \mathcal{N} \delta^2$ of the portal coupling. In case the BSM sector also develops a vacuum-expectation value, mixing between the SM and the BSM Higgs modes arise, leading to modifications of the triple- and quartic Higgs vertices that are testable at future colliders~\cite{Hiller:2024zjp}.

Finally, there are also non-minimal models which leverage a blend of these portals. Non-minimal extensions might also exhibit other mechanisms such as supersymmetry. 

\section{Conclusion}
Using state-of-the-art theoretical tools and experimental input, we have updated the stability analysis of the Higgs potential in the SM. Evidence suggests that the SM is in a metastable regime. However, the scenario of absolute stability cannot be excluded decisively. To do so, the experimental uncertainty of top mass and strong coupling constant need to be reduced and their correlation taken into account. As for the top, quantifying the relation and theory uncertainty between the physical and Monte Carlo mass might offer a significant improvement of precision.

We have presented several pathways to stabilise the Higgs potential via SM extensions. They lead to minimal models with new physics as low as the TeV range and signatures testable at current colliders. Furthermore, such BSM theories are valid for a vast energy range up to the Planck scale, rendering them candidates for robust and predictive extensions of the SM rather than a short-lived effective field theories. In that sense, they may account for \textit{great desert} scenarios of no additional new physics up to the Planck scale.

\section*{Acknowledgements}
These proceedings are based on an invited plenary talk at the 17th International Workshop on Top Quark Physics (Top2024). I am grateful to the organisers for their kind invitation and the participants for stimulating discussions. I am also thankful to my collaborators G.~Hiller, T.~H\"ohne and D.~F.~Litim for comments on this manuscript.

\bibliography{ref.bib}

\end{document}